%% file: main.tex
\documentclass{aastex701}

\graphicspath{{./}{figures/}}

\usepackage{amsmath}
\usepackage{bm}
\usepackage{color}
\usepackage{hyperref}
\usepackage{enumitem}

\newcommand{\psr}{PSR~J0248+6021}
\newcommand{\frb}{FRB~20191221A}

\begin{document}

\title{A series of unfortunate events: CHIME/FRB misclassification of a Galactic pulsar as a periodic fast radio burst}

\include{auth}

\begin{abstract}
In 2022, the CHIME/FRB Collaboration reported the detection of FRB 20191221A, an apparent fast radio burst exhibiting a significant periodicity of 217\,ms.
Recently, this event has been identified as a series of pulses originating from the known Galactic pulsar PSR~J0248$+$6021.
The initial misidentification was caused by an unusual calibration problem with the CHIME telescope, coupled with the atypical characteristics of the pulsar's emission. 
Here, we detail the issues with the calibration and how it led to a many-degree offset in the pointing of the calculated beams.
We describe how we verified that this problem has not affected any other FRB localization, including those reported in the Second CHIME/FRB Catalog, and our newly implemented checks to ensure this mispointing problem does not affect future data.
\end{abstract}


\section{The detection of a unique signal}
The fast radio burst (FRB) backend of the Canadian Hydrogen Intensity Mapping Experiment (CHIME) detected an unusual signal on 2019 December 21.
Its dispersion measure (DM) exceeded the maximum predictions of Galactic models in what at the time was believed to be its direction by a factor of four \citep{ne2001,ymw2016}, suggesting an extragalactic nature consistent with that of FRBs \citep{lorimer+2007}.
The event, labelled as \frb, was unique among the FRB population due to its long duration ($\sim 3$\,s) and nine significant peaks in its pulse profile. 
A timing analysis revealed a $>6\sigma$ detection of a $\sim 217$\,ms periodicity between the times of arrival of those peaks, making it the only known FRB with a significant periodicity detected in its signal, as reported by \citet{chimefrb_subsecperiodicity}.
This value of the period is typical for the Galactic pulsar population.
This event also had one of the largest scattering timescales measured for an FRB, suggesting a turbulent and, possibly, dense environment, which could have caused the DM excess and implied a Galactic source. 

\citet{chimefrb_subsecperiodicity} reported the discovery of \frb\ and its periodicity.
In the article, the similarities with signals from Galactic pulsars were recognized. 
However, the large DM excess at the location of the event pointed to an extragalactic origin, reinforced by the lack of ionized or star-forming regions in the localization region \citep{chimefrb_subsecperiodicity}. 
As detailed below, we later discovered a mispointing of the digital beams of the telescope, which caused a wrong event position, rendering the measured DM excess unreliable.
In fact, the event we called \frb\ was actually produced by a few pulses from the high-DM Galactic pulsar \psr\ detected at an incorrect location.
After discovering the issue, we promptly informed the community during the FRB~2025 conference in Montreal \citep{curtin+2026_frb2025}, and are in the process of retracting the article that presented \frb\ \citep{chimefrb_subsecperiodicity}.

\section{Something was wrong}
Two circumstances prompted the collaboration to investigate the \frb\ data.
First, CHIME/Slow \citep{mat+2026_chimeslow}, the new backend of CHIME designed to discover longer-duration FRBs, detected events from \psr\ that show a striking resemblance to the morphology of \frb. Further analysis revealed that the period, dispersion measure (DM), and right ascension (RA) of events from the two sources were compatible, although the declinations (Dec) were roughly $20^\circ$ discrepant.
Second, independent work on the Second CHIME/FRB Baseband Catalog identified two bursts (labeled \emph{twin bursts} for their similarity) detected on the same day as \frb, 
which had identical values of DM and time of arrival, and consistent RA and Faraday rotation, but were separated by approximately 20 degrees in declination.
These facts showed that \frb\ was likely a series of pulses from \psr\ detected at the wrong Dec.
Run notes for the experiment on the detection day and the raw voltage data stored for the twin bursts were then used to investigate the origin of incorrect Dec values, allowing us to identify an issue with the calibration of the telescope data.
We then verified that, when applying good gain solutions from a different day, the copy of the twin bursts at the incorrect location disappeared (Fig~\ref{fig:gains_offset}).

\section{The culprit: a corrupted calibration solution}
\begin{figure}
    \centering
    \includegraphics[width=\linewidth,keepaspectratio]{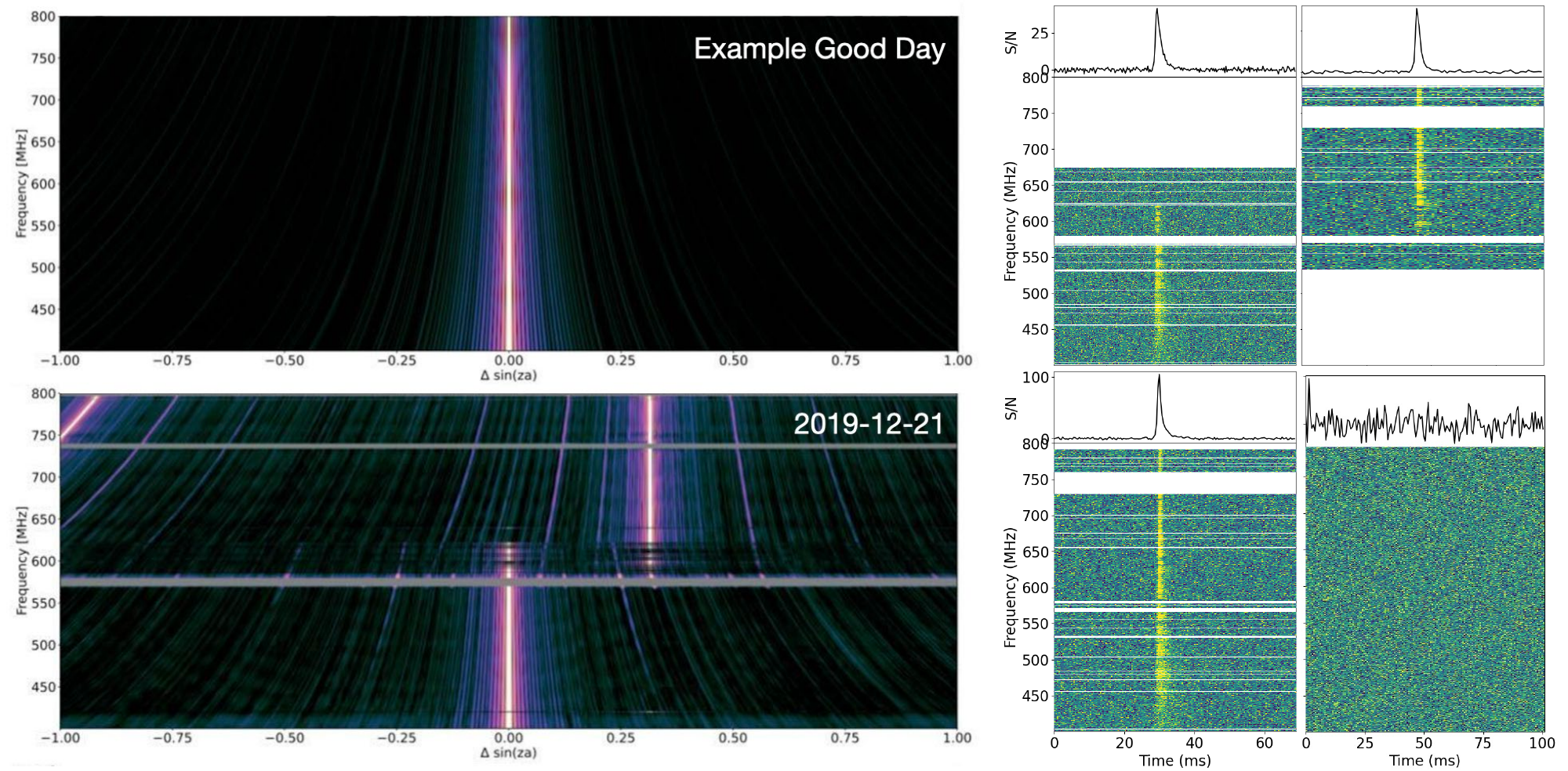}
    \caption{Left: Diagnostic derived from daily CHIME gain solutions for a representative good day (top) and for 2019 December 21 (bottom). For each day, the daily gains are compared to a set of historically good gains, the per-feed gain ratio is beamformed onto a two-dimensional grid spanning the CHIME field of view, and the result is collapsed by taking the maximum along the east--west direction-cosine axis $l$. The plotted quantity is therefore shown as a function of observing frequency and the orthogonal direction-cosine axis $m$ (approximately $\Delta \sin(\mathrm{za})$ near the meridian). On a good day, the peak remains centered near $m=0$ across the band. On 2019 December 21, the peak in the upper part of the band is offset by $\sim 0.30$ in direction cosine (roughly $20^\circ$ at zenith), indicating a large frequency-dependent pointing offset caused by the corrupted calibration.
    Right: Waterfall plots for FRB~20191221D, aka the \emph{twin bursts}, beamformed at Dec~$\sim42\deg$ (left) and Dec~$\sim60\deg$ (right) with bad gains (top) and good gains (bottom). While the bad gains yield a ghost copy of the burst at the frequency channels affected by calibration issues, this ghost disappears when using good gains, and the full signal is recovered at the right position.
    }
    \label{fig:gains_offset}
\end{figure}

On 2019 December 21, heavy rain caused a subset of feeds to malfunction. This is consistent with a known failure mode in which water pools in the focal line and the electronics become wet, resulting in increased noise, reduced gain, and occasional amplifier oscillations, with effects typically strongest in the upper portion of the 400--800 MHz band.
CHIME determines per-feed complex gain solutions once per day from observations of a bright transiting calibrator (typically Cygnus~A, Cassiopeia~A, or Taurus~A) using a visibility-based data stream separate from the main CHIME/FRB search path \citep{chime_sys_overview}. At the time, a newly deployed RFI-excision algorithm was being tested on this alternate data stream. The algorithm used a feed-averaged spectral-kurtosis metric, and the wet feeds caused outlier values of this metric, resulting in the upper part of the band being flagged as bad for several hours, including during the calibrator transit used to compute that day’s gain solution.
Our investigation indicates that, in this edge case where all data during the calibrator transit was flagged, the calibration pipeline likely used stale visibility data rather than the intended calibrator-transit data.  Although the exact software failure mode could not be fully reconstructed, the resulting gain solution is consistent with calibration from mismatched source data and exhibits a large phase gradient as a function of north--south feed position in the upper part of the band ($\gtrsim 600$\,MHz). This is expected if the solver assumes one source position while the flagged frequencies are instead using stale data dominated by a different source, since the inferred gains then absorb the phase difference corresponding to the angular separation between the intended and actual calibration sources.
When applied in beamforming, this phase gradient produced a substantial declination pointing offset at those frequencies, while lower frequencies remained correctly pointed.
This effect is clearly illustrated in Fig.~\ref{fig:gains_offset}. The left plots show a diagnostic constructed from the gain solutions on both a good day and the day in question. 

Within a few hours, the CHIME team identified the gain solution to be of poor quality based on a large discontinuity in the feed-averaged autocorrelations. CHIME/FRB then reverted to a previous good gain solution. However, the specific failure mode --- a systematic phase gradient as a function of feed position in only the upper portion of the band, causing a frequency-dependent pointing offset --- was not recognized at the time. Consequently, events detected in the few hours between the application of the bad gains and their reversion were not flagged by CHIME/FRB.
Furthermore, CHIME/FRB uses a large sample of pulses detected from many known pulsars to continuously check the quality of its data. 
Normally, at CHIME observing frequencies, pulsars are brighter at lower frequencies, which were unaffected by the bad calibration, thus showing a healthy system. 
However, probably due to the large scattering experienced by \psr, CHIME detected it at the wrong beam positions.
Upon closer inspection, a small fraction of faint bursts from the bright PSR~B0329+54 were also found to be offset by about $20^\circ$ in Dec.

Coincidentally, \psr\ has an abnormally large DM from a dense \ion{H}{2} region \citep{theu+2011}, and it is located $\sim 0.7$ degrees from the Galactic plane. 
This caused it to be classified as likely extragalactic by CHIME/FRB pipelines at the wrong offset of $\sim 20$ degrees from its true position. 
Moreover, \psr\ has a peculiar emission pattern similar to those of rotating radio transients \citep{wel+2010_rrats}, with bright bursts of activity, which further strengthened its similarities with FRBs and prevented us from recognising it as a Galactic pulsar. 
Finally, during manuscript preparation, both \ion{H}{2} region catalogs \citep{and+2014_hii,gre2019_snrs} and the ATNF pulsar catalog\footnote{\url{http://www.atnf.csiro.au/research/pulsar/psrcat}} \citep{man+2005_atnf} were checked for possible evidence of a Galactic origin of the event, but not up to a 20-degree offset in Dec as there was no systematic with the CHIME/FRB instrument known at the time that could lead to such a drastic offset.

\section{Verification of past and future detections} 
To exclude the possibility that the same issue affects other CHIME/FRB detections, a diagnostic was developed to verify the quality of the daily calibration used to calculate CHIME beam pointings.
Since this particular issue manifests as a shift in the gains at some frequencies, for each daily calibration, we form a grid of beams spanning the primary beam, integrate over frequency, and search for peaks at non-zero offset.  
For any calibration where we detect a second peak exceeding a threshold of 1\% of the primary peak at zero offset, we visually inspect the beam grid resolved in frequency (as in Fig.~\ref{fig:gains_offset}).
For past observations, we found a secondary peak exceeding the threshold in about $1\%$ of daily gains.
In most of these cases, the shift was observed within a narrow frequency band (less than $10$\,MHz), likely due to narrowband RFI during the calibrator transit that dominates the signal in a limited frequency range, causing the gain solution to pick up a positional offset corresponding to the separation between the RFI source and the assumed calibrator position.  The frequency dependence of these offsets is also consistent with distant TV stations scattered by aircraft, satellites, or meteor ionization trails.
For every event detected on those days, we verified that the spectral flux density was not dominated by signal in these narrow frequency ranges.
Only two days showed a large number of frequency channels that exhibited a significant shift: 2019 December 21, the day we detected \frb, and 2019 March 11, which did not have FRB detections.
We note that FRB~20191221B, also discovered on 2019 December 21 and reported by \citet{chimefrbcatalog2}, was detected after the gains were reversed to the good solutions obtained on the previous day.
We have applied this check to all FRBs published by CHIME/FRB, including the recently released Second CHIME/FRB Catalog \citep{chimefrbcatalog2}, and we will apply this check to all detected CHIME FRBs.

\section{Acknowledgements}
We acknowledge that CHIME is located on the traditional, ancestral, and unceded territory of the Syilx/Okanagan people. We are grateful for the warm reception and skillful help we have received from the staff of the Dominion Radio Astrophysical Observatory, which is operated by the Nati
onal Research Council of Canada.

CHIME operations are funded by a grant from the NSERC Alliance Program and by support from McGill University, University of British Columbia, and University of Toronto. CHIME was funded by a grant from the Canada Foundation for Innovation (CFI) 2012 Leading Edge Fund (Project 31170) and by contributions from the provinces of British Columbia, Québec and Ontario. The CHIME/FRB Project was funded by a grant from the CFI 2015 Innovation Fund (Project 33213) and by contributions from the provinces of British Columbia and Québec, and by the Dunlap Institute for Astronomy and Astrophysics at the University of Toronto. Additional support was provided by the Canadian Institute for Advanced Research (CIFAR), the Trottier Space Institute at McGill University, and the University of British Columbia.

\allacks{}

\bibliography{refs}{}
\bibliographystyle{aasjournalv7}

\end{document}

%% file: auth.tex
\collaboration{all}{The CHIME/FRB Collaboration}

\author[0000-0001-5908-3152]{Bridget C.~Andersen}
  \affiliation{Department of Physics, McGill University, 3600 rue University, Montr\'eal, QC H3A 2T8, Canada}
  \affiliation{Trottier Space Institute, McGill University, 3550 rue University, Montr\'eal, QC H3A 2A7, Canada}
  \affiliation{Department of Astronomy and Astrophysics, University of California, Santa Cruz, 1156 High Street, Santa Cruz, CA 95060, USA}
  \email{}
\author[0000-0002-3615-3514]{Mohit Bhardwaj}
  \affiliation{Department of Space, Planetary \& Astronomical Sciences \& Engineering, Indian Institute of Technology Kanpur, Helicopter Building, Kalyanpur, Kanpur, Uttar Pradesh 208016}
  \email{}
\author[0000-0001-8537-9299]{P.~J.~Boyle}
  \affiliation{Department of Physics, McGill University, 3600 rue University, Montr\'eal, QC H3A 2T8, Canada}
  \email{}
\author[0000-0003-2047-5276]{Tomas Cassanelli}
  \affiliation{Department of Electrical Engineering, Universidad de Chile, Av. Tupper 2007, Santiago 8370451, Chile}
  \email{}
\author[0000-0002-2878-1502]{Shami Chatterjee}
  \affiliation{Cornell Center for Astrophysics and Planetary Science, Cornell University, Ithaca, NY 14853, USA}
  \email{}
\author[0000-0003-2319-9676]{Davor Cubranic}
  \affiliation{Department of Physics and Astronomy, University of British Columbia, 6224 Agricultural Road, Vancouver, BC V6T 1Z1 Canada}
  \email{}
\author[0000-0001-8384-5049]{Emmanuel Fonseca}
  \affiliation{Department of Physics and Astronomy, West Virginia University, PO Box 6315, Morgantown, WV 26506, USA }
  \affiliation{Center for Gravitational Waves and Cosmology, West Virginia University, Chestnut Ridge Research Building, Morgantown, WV 26505, USA}
  \email{}
\author[0000-0002-3382-9558]{B.~M.~Gaensler}
  \affiliation{Department of Astronomy and Astrophysics, University of California, Santa Cruz, 1156 High Street, Santa Cruz, CA 95060, USA}
  \affiliation{Dunlap Institute for Astronomy and Astrophysics, 50 St. George Street, University of Toronto, ON M5S 3H4, Canada}
  \affiliation{David A.\ Dunlap Department of Astronomy and Astrophysics, 50 St. George Street, University of Toronto, ON M5S 3H4, Canada}
  \email{}
\author[0000-0003-1884-348X]{Deborah C.~Good}
  \affiliation{Department of Physics and Astronomy, University of Montana, 32 Campus Drive, Missoula, MT 59812, USA}
  \email{}
\author[0000-0003-4810-7803]{Jane F.~Kaczmarek}
  \affiliation{SKA Observatory, Science Operations Centre, 26 Dick Perry Avenue, Kensington WA 6151 Australia }
  \email{}
\author[0000-0002-3354-3859]{Joseph Kania}
  \affiliation{Jodrell Bank Centre for Astrophysics, University of Manchester, Alan Turing Building, Oxford Road, Manchester, M13 9PL}
  \email{}
\author[0000-0001-9345-0307]{Victoria M.~Kaspi}
  \affiliation{Department of Physics, McGill University, 3600 rue University, Montr\'eal, QC H3A 2T8, Canada}
  \affiliation{Trottier Space Institute, McGill University, 3550 rue University, Montr\'eal, QC H3A 2A7, Canada}
  \email{}
\author[0000-0002-4279-6946]{Kiyoshi W.~Masui}
  \affiliation{MIT Kavli Institute for Astrophysics and Space Research, Massachusetts Institute of Technology, 77 Massachusetts Ave, Cambridge, MA 02139, USA}
  \affiliation{Department of Physics, Massachusetts Institute of Technology, 77 Massachusetts Ave, Cambridge, MA 02139, USA}
  \email{}
\author[0000-0002-2551-7554]{Daniele Michilli}
  \affiliation{Laboratoire d'Astrophysique de Marseille, Aix-Marseille Univ., CNRS, CNES, Marseille, France}
  \email[show]{danielemichilli@gmail.com}
\author[0000-0002-7333-5552]{Laura Newburgh}
  \affiliation{Department of Physics, Yale University, New Haven, CT 06520, USA}
  \email{}
\author[0000-0002-2465-8937]{Anna Ordog}
  \affiliation{Department of Physics \& Astronomy, University of Western Ontario, 1151 Richmond Street, London, ON, N6A 3K7, Canada}
  \email{}
\author[0000-0002-8912-0732]{Aaron B.~Pearlman}
  \altaffiliation{NASA Hubble Fellow}
  \affiliation{MIT Kavli Institute for Astrophysics and Space Research, Massachusetts Institute of Technology, 77 Massachusetts Ave, Cambridge, MA 02139, USA}
  \affiliation{Department of Physics, Massachusetts Institute of Technology, 77 Massachusetts Ave, Cambridge, MA 02139, USA}
  \affiliation{Department of Physics, McGill University, 3600 rue University, Montr\'eal, QC H3A 2T8, Canada}
  \affiliation{Trottier Space Institute, McGill University, 3550 rue University, Montr\'eal, QC H3A 2A7, Canada}
  \email{}
\author[0000-0003-2155-9578]{Ue-Li Pen}
  \affiliation{Institute of Astronomy and Astrophysics, Academia Sinica, Astronomy-Mathematics Building, No. 1, Sec. 4, Roosevelt Road, Taipei 106319, Taiwan}
  \affiliation{Canadian Institute for Theoretical Astrophysics, 60 St.~George Street, Toronto, ON M5S 3H8, Canada}
  \affiliation{Canadian Institute for Advanced Research,  661 University Avenue, Toronto, Ontario M5G 1M1, Canada}
  \affiliation{Dunlap Institute for Astronomy and Astrophysics, 50 St. George Street, University of Toronto, ON M5S 3H4, Canada}
  \affiliation{Perimeter Institute of Theoretical Physics, 31 Caroline Street North, Waterloo, ON N2L 2Y5, Canada}
  \email{}
\author[0000-0002-9822-8008]{Emily Petroff}
  \affiliation{Perimeter Institute of Theoretical Physics, 31 Caroline Street North, Waterloo, ON N2L 2Y5, Canada}
  \email{}
\author[0000-0002-4795-697X]{Ziggy Pleunis}
  \affiliation{Anton Pannekoek Institute for Astronomy, University of Amsterdam, Science Park 904, 1098 XH Amsterdam, The Netherlands}
  \affiliation{ASTRON, Netherlands Institute for Radio Astronomy, Oude Hoogeveensedijk 4, 7991 PD Dwingeloo, The Netherlands}
  \email{}
\author[0000-0003-1842-6096]{Mubdi Rahman}
  \affiliation{Sidrat Research, 124 Merton Street, Suite 507, Toronto, Ontario, M4S 2Z2, Canada}
  \email{}
\author[0000-0001-5799-9714]{Scott Ransom}
  \affiliation{National Radio Astronomy Observatory, 520 Edgemont Rd, Charlottesville, VA 22903, USA}
  \email{}
\author[0000-0003-3463-7918]{Andre Renard}
  \affiliation{Dunlap Institute for Astronomy and Astrophysics, 50 St. George Street, University of Toronto, ON M5S 3H4, Canada}
  \email{}
\author[0000-0001-5504-229X]{Pranav Sanghavi}
  \affiliation{Center for Astrophysics | Harvard \& Smithsonian, 60 Garden Street, Cambridge, MA 02138, USA}
  \email{}
\author[0000-0002-7374-7119]{Paul Scholz}
  \affiliation{Department of Physics and Astronomy, York University, 4700 Keele Street, Toronto, ON MJ3 1P3, Canada}
  \email{}
\author[0000-0002-6823-2073]{Kaitlyn Shin}
  \affiliation{Cahill Center for Astronomy and Astrophysics, MC 249-17 California Institute of Technology, Pasadena CA 91125, USA}
  \email{}
\author[0000-0003-2631-6217]{Seth R.~Siegel}
  \affiliation{SKA Observatory, Science Operations Centre, 26 Dick Perry Avenue, Kensington WA 6151 Australia }
  \affiliation{Perimeter Institute of Theoretical Physics, 31 Caroline Street North, Waterloo, ON N2L 2Y5, Canada}
  \affiliation{Department of Physics, McGill University, 3600 rue University, Montr\'eal, QC H3A 2T8, Canada}
  \affiliation{Trottier Space Institute, McGill University, 3550 rue University, Montr\'eal, QC H3A 2A7, Canada}
  \email{}
\author[0000-0001-7755-902X]{Saurabh Singh}
  \affiliation{Raman Research Institute, C. V. Raman Avenue, Sadashivanagar, Bangalore, Karnataka - 560080, India}
  \email{}
\author[0000-0002-2088-3125]{Kendrick M.~Smith}
  \affiliation{Perimeter Institute of Theoretical Physics, 31 Caroline Street North, Waterloo, ON N2L 2Y5, Canada}
  \email{}
\author[0000-0003-4535-9378]{Keith Vanderlinde}
  \affiliation{Dunlap Institute for Astronomy and Astrophysics, 50 St. George Street, University of Toronto, ON M5S 3H4, Canada}
  \affiliation{David A.\ Dunlap Department of Astronomy and Astrophysics, 50 St. George Street, University of Toronto, ON M5S 3H4, Canada}
  \email{}
\author[0000-0001-7314-9496]{Dallas Wulf}
  \affiliation{Department of Physics, McGill University, 3600 rue University, Montr\'eal, QC H3A 2T8, Canada}
  \affiliation{Trottier Space Institute, McGill University, 3550 rue University, Montr\'eal, QC H3A 2A7, Canada}
  \email{}
\author[0000-0001-8278-1936]{Andrew V.~Zwaniga}
  \affiliation{Toronto Metropolitan University, 350 Victoria Street, Toronto, ON M5B 2K3, Ontario, Canada
  \\\\
  }
  \email{}
\newcommand{\allacks}{
The Dunlap Institute is funded through an endowment established by the David Dunlap family and the University of Toronto.
Research at Perimeter Institute is supported in part by the Government of Canada through the Department of Innovation, Science and Economic Development and by the Province of Ontario through the Ministry of Colleges, Universities, Research Excellence and Security.
The AstroFlash research group at McGill University, University of Amsterdam, ASTRON, and JIVE is supported by: a Canada Excellence Research Chair in Transient Astrophysics (CERC-2022-00009); an Advanced Grant from the European Research Council (ERC) under the European Union’s Horizon 2020 research and innovation programme (`EuroFlash’; Grant agreement No. 101098079); an NWO-Vici grant (`AstroFlash’; VI.C.192.045); an NSERC Discovery Grant (RGPIN-2025-06681); an ERC Starting Grant (`EnviroFlash’; Grant agreement No. 101223057); and an NWO-Veni grant (VI.Veni.222.295).
B.\,C.\,A. was supported by an FRQNT Doctoral Research Award for most of this work.
E.F. is supported by the National Science Foundation under grant AST-2407399.
DCG is supported by NSF Astronomy and Astrophysics Grant (AAG) award 2406919.
V.M.K. holds the Lorne Trottier Chair in Astrophysics \& Cosmology, a Distinguished James McGill Professorship, and receives support from an NSERC Discovery grant (RGPIN 228738-13).
K.W.M. is supported by NSF Grant No. 2510771 and receives lumbar support from the Adam J. Burgasser Chair in Astrophysics.
D.M. acknowledges support from the French government under the France 2030 investment plan, as part of the Initiative d'Excellence d'Aix-Marseille Universit\'e -- A*MIDEX (AMX-23-CEI-088).
L.N. receives support from NSF Grant No. 2510772
A.B.P. acknowledges support by NASA through the NASA Hubble Fellowship grant HST-HF2-51584.001-A awarded by the Space Telescope Science Institute, which is operated by the Association of Universities for Research in Astronomy, Inc., under NASA contract NAS5-26555. A.B.P. also acknowledges prior support from a Banting Fellowship, a McGill Space Institute~(MSI) Fellowship, and a Fonds de Recherche du Quebec -- Nature et Technologies~(FRQNT) Postdoctoral Fellowship.
U.P. is supported by the Natural Sciences and Engineering Research Council of Canada (NSERC) [funding reference number RGPIN-2019-06770, ALLRP 586559-23, RGPIN-2025-06396], Canadian Institute for Advanced Research (CIFAR), Ontario Research Fund – Research Excellence (ORF-RE Fund, 72074697), AMD AI Quantum Astro. Thoth Technology Inc. Big Sky (Fund number 522968)
SMR is a CIFAR Fellow and is supported by the NSF Physics Frontiers Center award 2020265.
P.S. acknowledges the support of an NSERC Discovery Grant (RGPIN-2024-06266).
}